\documentclass[twocolumn,superscriptaddress,showpacs,aps,floatfix]{revtex4-1}
\usepackage{amsmath,amsfonts,amssymb,epsfig,dcolumn,bm,dsfont,graphics,latexsym,color,graphicx}
\def\ri{\mathrm{i}}

\let\oldAA\AA

\def\ri{\mathrm{i}}
\renewcommand{\AA}{\text{\normalfont\oldAA}}
\newcommand{\ket}[1]{| {#1} \rangle} 
\newcommand{\bra}[1]{\langle {#1} |} 
\newcommand{\braket}[2]{\langle {#1} \vphantom{#2} | {#2} \vphantom{#1} \rangle} 
\begin{document}
\preprint{AIP/123-QED}
\title{Electrical control of the hole spin qubit in Si and Ge nanowire quantum dots}
\author{Marko Milivojevi\'c}
\affiliation {NanoLab, QTP Center, Faculty of Physics, University of Belgrade, Studentski trg 12, 11001 Belgrade, Serbia}
\affiliation{Institute of Physics, Pavol Jozef \v{S}af\'{a}rik University, Park Angelinum 9, 040 01 Ko\v{s}ice, Slovak Republic}
\begin{abstract}
Strong, direct Rashba spin-orbit coupling in Si, Ge, and the Ge/Si core/shell nanowire quantum dot (QD) allows for all electrical manipulation of the hole spin qubit. 
Motivated by this fact, we analyze different fabrication-dependent properties of nanowires, such as orientation, cross section, and the presence of strain, with the goal being to find the
material and geometry that enables
the fastest qubit manipulation, whose speed can be identified using the Rabi frequency. 
We show that QD in nanowires with a circular cross section (cNWs) enables much weaker driving of the hole spin qubit than QDs embedded in square profile nanowires (sNWs).
Assuming the orientation of the Si nanowire that maximizes the spin-orbit effects, our calculations predict that the Rabi frequencies of the hole spin qubits inside Ge and Si sNW QD have comparable strengths for weak electric fields. 
The global maximum of the Rabi frequency is found in Si sNW QD for strong electric fields,  putting this setup ahead of others in creating the hole spin qubit.
Finally, we demonstrate that strain in the Si/Ge core/shell nanowire QD decreases the Rabi frequency.
In cNW QD, this effect is weak; in sNW QD, it is possible to optimize the impact of strain with the appropriate tuning of the electric field strength.
\end{abstract}
\maketitle
\section{Introduction}
The electron or hole spin trapped inside a semiconductor quantum dot (QD) can be used as a building block of a quantum computer~\cite{NC10,BdV00}. 
To this end, approaches based on the magnetic~\cite{KBT+06,PLZ+08} and electric~\cite{GBL06,NKN+07} fields to manipulate the spin qubit are suggested. 
Even though the control of the spin qubit is
more straightforward with magnetic fields, electrical control of the spin qubit using the
electric-dipole spin resonance (EDSR) is favorable in physical realizations~\cite{KSW+14,TYO+18,KGS12,KLS19,KLS20,SM20,MSD+20,BBA+21}.

The principal physical mechanism that enables the electrical control of the spin qubit is the
spin-orbit interaction (SOC). Besides its positive effect in EDSR-based schemes, SOC leads to undesirable effects such as decoherence and relaxation~\cite{GKL04,SF06,MKL13,LVN20}. Among materials with notable SOC that can host spin qubits, Si and Ge have recently attracted much attention due to their free nuclear spin environment, leading to long dephasing times~\cite{TTL+12,VHY+14,VYH+15}. Also, in the group of different Si and Ge nanostructures, special interest is devoted to quasi-one-dimensional geometries such as hut wires~\cite{ZKM12,WKV+16,WKL+18,GWW+20,XLG+20}
and nanowires~\cite{HKL+12,HLY+14,BRL+16,KRL18,FRR+18,FRH+21}. In such systems, the realization of spin qubits with holes rather than with electrons is owed to the fact that SOC is much stronger in the valence than in the  conduction bands. The so-called "direct Rashba spin-orbit interaction" (DRSOI) that was predicted in Ge/Si core/shell nanowires~\cite{KTL11} represents an efficient way to manipulate hole spin states in such structures electrically
~\cite{HTC+10,HKL+14}. The other SOC mechanisms, Dresselhaus~\cite{DR}/Rashba~\cite{RA},
are forbidden by symmetry/much weaker than the DRSOI term.

Here we investigate how the electrical control of a hole spin qubit in Si, Ge, and Ge/Si core/shell nanowire QD is dependent on the electrically tunable DRSOI and fabrication-dependent parameters of the nanowire, such as orientation, cross section, and strain. We focus on realistic profile shapes, circular and square cross sections~\cite{BCC+12,PMB+12}, and realistic profile sizes~\cite{VMB+16}. 
Since the hole states in Ge have almost isotropic dispersion relation at the $\Gamma$ point,
we employed the spherical approximation when studying the hole spin qubit in Ge and Ge/Si core/shell nanowire QD. On the other hand, the orientation dependence of Si hole states is taken into account when discussing the hole spin qubit in Si nanowire QD. 

The Rabi frequency, measuring the speed of single-qubit rotations, can be used to assess the efficiency  of the hole spin qubit. We used the fact that the strongest $g$ factor is achieved when the electric and magnetic fields are applied perpendicular to the nanowire and are mutually parallel~\cite{KRL18}; the strong driving of the Rabi frequency is enabled by varying the electric field strength. 
We divided the Rabi frequency dependence on the electric field strength into two regimes: in the first one, Rabi frequency is proportional to the electric field strength, while in the second regime a nonlinear response is observed.
Our analysis shows that hole spin qubits in circular cross section nanowire (cNW) QDs
are much less efficiently controlled by the electric field than in square profile nanowire (sNW) QDs.
In the linear regime, we showed that the Rabi frequency in hole spin qubits inside Ge and Si sNW QD are of comparable strength, assuming the orientation of Si nanowire such that the spin-orbit effects are maximized. 
In the nonlinear regime, the global maximum of Rabi frequency is found in Si sNW QD, putting this setup in favor of others for the creation of the hole spin qubit.
We also investigated the role of strain in the hole spin qubit formed in the Ge/Si core/shell nanowire QD. We have shown that strain always decreases the Rabi frequency;
in cNW QDs this effect is not so pronounced, whereas in sNW QDs the strong impact of shell thickness can be minimized with the appropriate tuning of the electric field strength.

This paper is organized as follows. After the introductory section, in Sec.~\ref{Hole qubit model} the model of the hole spin qubit in Si, Ge, and Ge/Si core/shell nanowire QD is introduced. The numerical procedure used for Hamiltonian diagonalization and the formal definition of the Rabi frequency is given in the same section. 
In Secs.~\ref{SiEDSRsection} and~\ref{GeEDSRsection} the dependence of the Rabi frequency on the electric field strength in Si and Ge cNW and sNW QD is analyzed, respectively. Next, in Sec.~\ref{StrainSection}, the role of strain in Ge/Si core/shell nanowire QD hole spin qubit on the Rabi frequency is discussed. Finally, in Sec.~\ref{Conclusions}, short conclusions are given.

\section{Model}\label{Hole qubit model}
\begin{figure}
\centering
\includegraphics[width=8.5cm]{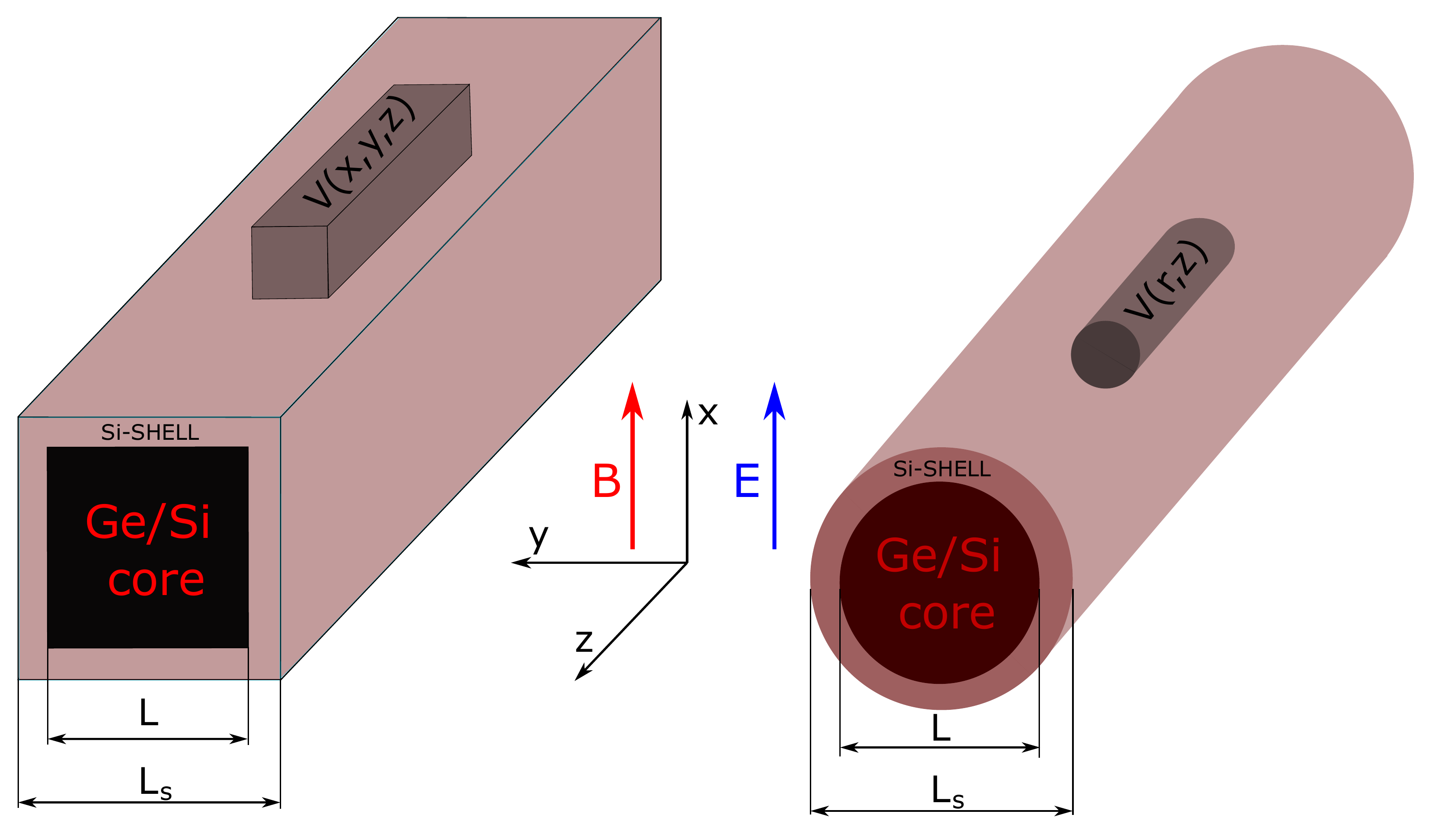}
\caption{{\it Setup for the construction of the hole spin qubit}.
The main axis of the nanowire coincides with the $z$-direction.
The core of the nanowire has a square (left panel) or circular (right panel) cross section.
In the case of the Ge/Si core/shell nanowire, the Si shell
around the Ge core increases the overall cross section,
where the shell thickness parameter $\gamma=(L_S-L)/L$ is given with the help of the outer Si shell diameter/side length $L_S$ 
and the inner Ge core diameter/side length $L$.
Note that no shell is present in the case of the Si and Ge nanowire ($\gamma=0$).
Gating potential $V$ used to localize the hole spin qubit, given in Eqs.~\ref{squarePOT} and \ref{circPOT}, is also denoted. Finally, both the magnetic and electric fields are applied in the $x$-direction to maximize the spin-orbit effects.}
\label{NWsetup}
\end{figure}

In this section, we introduce the setup for the creation of the hole spin qubit (see FIG.~\ref{NWsetup} for an illustration), whose dynamics can be described using the Hamiltonian
\begin{equation}\label{hole-model}
H=H_{\rm LK}+H_{\rm DRSOI}+H_{\rm Z}+V+H_{\rm BP}.
\end{equation}
The first term of the total Hamiltonian $H$ is the
four-band Luttinger-Kohn (LK) Hamiltonian~\cite{LK55,L56}.
In the simplest case, where the nanowire axes coincide with the crystallographic directions $[100]$, $[010]$, and $[001]$ ($xyz$ orientation), the LK Hamiltonian is equal to
\begin{eqnarray}\label{LKham}
  H_{\rm LK}^{\rm xyz}&=&\frac{\hbar^2}{2m}\big[(\gamma_1+\frac{5}2\gamma_2){\bf k}^2-2\gamma_2(k_x^2J_x^2+k_y^2J_y^2+k_z^2J_z^2)\nonumber\\
  &&-4\gamma_3(\{k_x,k_y\}\{J_x,J_y\}+c.p.)\big],
\end{eqnarray}
where c.p. stands for cyclic permutation, $\{A,B\}=(AB+BA)/2$ is the anticommutator,
$m$ is free electron mass, $\gamma_{1,2,3}$ are the Luttinger parameters,
${\bf k}$ is the momentum operator, and $J_i$ are spin-3/2 operators
obeying the relation $[J_a,J_b]=\ri\epsilon_{abc} J_c$
($\epsilon_{abc}$ is the Levi-Civita symbol).
Note that in zero magnetic field, the momentum operator is equal to
$-\ri\nabla$, while for homogeneous magnetic field
${\bf B}=B_x{\bf e}_x+B_y{\bf e}_y+B_z{\bf e}_z$, it
equals to ${\bf k}=-\ri\nabla+e/\hbar {\bf A}$ ($e>0$), where
the vector potential ${\bf A}$ is given as
\begin{equation}\label{vectorpot}
  {\bf A}=-\frac{1}2B_z y{\bf e}_x+\frac{1}2B_zx{\bf e}_y+(B_xy-B_yx){\bf e}_z.
\end{equation}
It is argued that the strongest spin-orbit effects~\cite{KRL18} occur when the
nanowire main axis is oriented along the [001] direction, whereas the $x$ and $y$ axes coincide with the [110] and [$\bar{1}10$] directions, respectively. In this case, the LK Hamiltonian is equal to~\cite{KRL18}
\begin{eqnarray}\label{LKhamROT}
   H_{\rm LK}^{\rm rot}&=&\frac{\hbar^2}{2m}
    \Big[(\gamma_1+\frac{5\gamma_2}2){\bf k}^2-\gamma_3(k_x^2-k_y^2)(J_x^2-J_y^2)\nonumber\\
    &&-4\gamma_3(\{k_y,k_z\}\{J_y,J_z\}+\{k_z,k_x\}\{J_z,J_x\})\nonumber\\
    &&-\gamma_2(k_x^2J_y^2+k_y^2J_x^2+4\{k_x,k_y\}\{J_x,J_y\})\nonumber\\
     &&-\gamma_2(k_x^2J_x^2+k_y^2J_y^2+2k_z^2J_z^2)\Big].
\end{eqnarray}
In the case of the Ge nanowire, the Luttinger parameters are equal to $\gamma_1 = 13.35$, $\gamma_2 = 4.25$, and $\gamma_3 = 5.69$~\cite{L71}, while the 
same parameters for the Si nanowire equal to $\gamma_1 = 4.22$, $\gamma_2 = 0.39$, and $\gamma_3 = 1.44$~\cite{L71}.
For Ge, the spherical approximation is valid since $\gamma_3/\gamma_2\approx1$.
Thus, instead of Hamiltonians~\eqref{LKham} and~\eqref{LKhamROT}
we will consider the LK Hamiltonian 
\begin{eqnarray}\label{LKhamSPHERICAL}
   H_{\rm LK}^{\rm spherical}&=&\frac{\hbar^2}{2m}
    \Big[(\gamma_1+\frac{5\gamma_s}2){\bf k}^2-2\gamma_s({\bf k}\cdot{\bf J})^2\Big],
\end{eqnarray}
invariant under arbitrary rotations of the
nanowire coordinate system with respect to the crystallographic axes,
where $\gamma_s=(2\gamma_2+3\gamma_3)/5=5.114$.

The second term in Eq.~\ref{hole-model} corresponds to the electric field-induced Hamiltonian
\begin{equation}\label{HDR}
H_{\rm DRSOI}=-e{\bf E}\cdot{\bf r}=-e(E_x x+E_y y+E_z z),
\end{equation}
usually called the DRSOI.
Our model neglects the Rashba SOC since it is shown that the DRSOI dominates~\cite{KRL18}. On the other hand, the Dresselhaus SOC is absent
in Ge and Si due to symmetry~\cite{W03}.

Furthermore, the direct coupling of the hole spin to the magnetic field
is described through the Zeeman term~\cite{W03}
\begin{equation}\label{hh}
  H_{\rm Z}=2k\mu_{B}{\bf B}\cdot{\bf J},
\end{equation}
with $\mu_B$ being the Bohr magneton, while $k=3.41$ (-0.26)~\cite{L71}
is the $g$ factor for Ge (Si) holes. Note that the anisotropic Zeeman term is
omitted, being a reasonable assumption in both materials~\cite{L71}.

Next, we describe the QD confinement potential $V$.
For a sNW the potential $V$ is equal to
\begin{equation}\label{squarePOT}
   V^{\rm s}= \left\{
      \begin{array}{rl}
      0, &  |x|<\frac{L}2, |y|<\frac{L}2, |z|<\frac{z_0}2\\
      \infty,&\,{\rm otherwise}.\\
      \end{array} \right.
\end{equation}
In the $xy$-plane, hard-wall conditions coincide with the
square profile of the nanowire, while the confinement strength
$z_0$ is dependent on the external gating.
In this work, we will assume $L=10$ nm~\cite{VMB+16}
and  $z_0=30$ nm~\cite{PPB+12}, being typical values
in nanowire qubit experiments.
Similarly, for a cNW, the gating potential is equal to
\begin{equation}\label{circPOT}
   V^{\rm c} = \left\{
      \begin{array}{rl}
      0, &  r<R=\frac{L}2, |z|<\frac{z_0}2\\
      \infty,&\,{\rm otherwise},\
      \end{array} \right.
\end{equation}
where $r$ represents the radial coordinate and $R=\frac{L}2$ is the half diameter of the profile.  

The last term in Eq.~\ref{hole-model} represents the strain-induced
Bir-Pikus Hamiltonian~\cite{BP74}, present only in the case of 
Ge/Si core/shell nanowire.
It is equal to
\begin{eqnarray}\label{BPham}
  &&H_{\rm BP}=-(a+\frac{5b}4)(\epsilon_{xx}+\epsilon_{yy}+\epsilon_{zz})\\
 &&+b(\epsilon_{xx}J_x^2+\epsilon_{yy}J_y^2+\epsilon_{zz}J_z^2)+\frac{2d}{\sqrt{3}}(\epsilon_{xy}\{J_x,J_y\}+c.p.),\nonumber
\end{eqnarray}
where $a$, $b$, $d$ are the deformation potentials, while $\epsilon_{ij}$ are the matrix elements of the symmetric strain tensor. In strained Ge/Si core/shell nanowires,
$\epsilon_{xy}=\epsilon_{xz}=\epsilon_{yz}=0$, $\epsilon_{xx}=\epsilon_{yy}=\epsilon_{\perp}$.
Furthermore, we use the following parameters: $a=2$eV, $b=-2.2$eV, $d=-4.4$eV~\cite{W70}. Since the hydrostatic deformation potential $a$ provides only a global shift that
can be discarded, the Bir-Pikus Hamiltonian has a very simple
effective form~\cite{KTL14}
\begin{equation}\label{BPeff}
  H_{\rm BP}^{\rm eff}=|b|(\epsilon_{\perp}(\gamma)-\epsilon_{zz}(\gamma))J_z^2.
\end{equation}
In the previous equation, functions $\epsilon_{\perp}$ and $\epsilon_{zz}$
depend on the parameter $\gamma=(L_S-L)/L$, described with the help of the outer Si shell diameter (side length) $L_S$ and the inner Ge core diameter (side length) $L$.
The dependence of $\epsilon_{\perp}$ and $\epsilon_{zz}$
on $\gamma$ and the general discussion of the validity of the model
can be found in~\cite{KTL14,KRL18}. 

As a final remark, it should be noted that there is a global minus sign in Eq.~\ref{hole-model} that does not affect the physics of holes. Its only effect is to resemble the most common positive (electron-like) energy levels of a particle in a box model and it is used in similar studies~\cite{KRL18}.
\subsection{Numerical diagonalization}\label{dijagonalizacija}
An adequate basis for the numerical diagonalization is needed to find the eigenvalues and eigenvectors of the hole spin qubit Hamiltonian $H$.
In the case of the hole spin qubit in sNW QD, 
the eigensolutions will be expanded in the basis set
\begin{equation}\label{eigen}
  \braket{\bf{r}}{n_xn_yn_z}\braket{s}{j_z}=\psi_{n_x}(x)\psi_{n_y}(y)\psi_{n_z}(z)\chi_{j_z},
\end{equation}
where
\begin{equation}\label{squaresolutions}
\psi_{n}(u)=\sqrt{\frac{2}{L_u}}\sin{\big(n\pi(\frac{u}{L_u}+\frac{1}2)\big)},
\end{equation}
$(n,u)=(n_x/n_y/n_z,x/y/z)$, represents solutions for the particle in the box model, with $n_{x,y,z}\geq 1$, $L_x=L_y=L$, and $L_z=z_0$; $\chi_{j_z}$ represents the eigenvector of the operator $J_z$,
\begin{equation}\label{spin}
  J_z\chi_{j_z}=j_z\chi_{j_z},
\end{equation}
with $j_z=\pm3/2,\pm1/2$.
On the other hand, in the case of the hole spin qubit in cNW QD, the eigenbasis is again adapted to the geometry of the problem and chosen as
\begin{equation}\label{eigen-circle}
  \braket{\bf{r}}{i n_z}
  \braket{s}{j_z}=\psi_{(m,n_r)}^{i}(r,\varphi)\psi_{n_z}(z)
  \chi_{j_z},
\end{equation}
where $\psi_{(m,n_r)}^{i}(r,\varphi)$ represents the ith eigenvector of the particle in an infinite circular well~\cite{R03}
\begin{equation}
\psi_{(m,n_r)}^i(r,\varphi)=N_{(m,n_r)}J_{(m,n_r)}(\frac{z_{m,n_r}}R r)\frac{{\rm e}^{\ri m\varphi}}{\sqrt{2\pi}},
\end{equation}
while $z_{m,n_r}$ is the $n_r$-th zero of the regular Bessel function $J_m(z)$ for $m=0,\pm1,\pm2,...$ quantized values of the angular momentum $L_z$. Additionally, $N_{(m,n_r)}$ represents the normalization constant, set by the equation $N_{(m,n_r)}^2\int_0^R J_{(m,n_r)}^2(\frac{z_{m,n_r}}R r)rdr=1$.

For sNW QD, numerical diagonalization is done using the 13500  basis states, i.e., the 15 lowest states in each Cartesian direction and four spin states.
On the other hand, for cNW QD, in addition to 15 $\psi_{n_z}(z)$ and four $\chi_{j_z}$ eigenstates, we used the 226 lowest $\psi_{(m,n_r)}(r,\varphi)$ eigenstates of the infinite circular well problem. We carefully checked that the given number of basis states enables good convergence of the Rabi frequency value for each configuration studied. This is done by comparing the Rabi frequency results with the $14\times14\times14\times4$-dimensional basis in the case of the square profile and the $196\times14\times4$-dimensional basis in the case of the circular profile. The estimated maximal relative difference is less than $0.35\%$ in the case of the square profile and less than $0.85\%$ for the circular profile, proving that the number of basis states is sufficient.
\subsection{Rabi frequency}\label{EDSR}
At zero magnetic field, the ground state of the Hamiltonian $H$
is two-fold degenerate. The presence of the magnetic field breaks the degeneracy and leads to the magnetic-field-dependent splitting between the initially (at $B=0$) degenerate ground state. We assume that qubit states $\ket{+}$ and $\ket{-}$ correspond to magnetic-field-induced split ground hole state at ${\bf B}=0$. With the applied oscillating electric field ${\bf E}_{\rm nw}$ in the direction of the nanowire main axis, we can achieve the electrical control of the hole spin qubit. When the oscillating field is resonant with the Larmor frequency of the qubit, the Rabi frequency, $\Omega_{\rm R}$, is equal to
\begin{equation}\label{RabiFreqDef}
  \Omega_{\rm R}=\frac{e}{\hbar}E_{\rm nw}|\bra{+}z\ket{-}|,
\end{equation}
where $E_{\rm nw}$ is the strength of the oscillating field. Since the strength of the Rabi frequency measures the speed of single-qubit rotations, the value of $\Omega_{\rm R}$ for different qubit configurations can be used to access the efficiency of the analyzed hole spin qubit.

The presence of mirror plane symmetries plays an essential role in determining the allowed directions of the electric and magnetic fields for obtaining $\Omega_{\rm R}\neq0$. 
Thus, we will shortly discuss their role.
First, in zero electric and magnetic field, the system is invariant under three mirror plane symmetries: $\sigma_{\rm xy}$, $\sigma_{\rm xz}$, and $\sigma_{\rm yz}$.
The applied oscillating electric field ${\bf E}_{\rm nw}$ in the $z$-direction breaks the
$\sigma_{\rm xy}$ symmetry, leaving only two mirror planes as symmetries of the system. If the static electric field is applied in the $z$-direction also, the $g$-matrix formalism~\cite{VBP+18} can be used to deduce that the Rabi frequency is zero, independently of the orientation of the magnetic field. On the other hand, the electric field applied in the $x$- or $y$-direction breaks the $\sigma_{\rm yz}/\sigma_{\rm xz}$ mirror plane, thus lowering the number of mirror plane-symmetries to one only. In this case,
the Rabi frequency is zero when the magnetic field is applied in the direction perpendicular to the mirror plane~\cite{VBP+18}, whereas in all other cases, nonzero $\Omega_{\rm R}$ is obtained. 

Since $\Omega_{\rm R}$ is dependent on the $g$ factor strength, it is plausible to set
the magnetic and electric field orientation in such a way that the value of $g$ is maximized.
This can be achieved when the magnetic field is perpendicular to the nanowire main axis and parallel to the electric field~\cite{KRL18}.
Thus, in this work we assume that both the
electric and magnetic fields are applied in the $x$-direction of the coordinate frame given in FIG.~\ref{NWsetup}.

\section{Hole spin qubit in Si nanowire QD}\label{SiEDSRsection}
We start our analysis of the Rabi frequency from hole spin qubits in the Si nanowire QD.
As discussed earlier, we will focus on the electric field effects, enabled by the presence of DRSOI.
For simplicity, we assume that the oscillating electric field strength ${\bf E}_{\rm nw}=0.03{\rm mV/nm}$ is fixed, as in~\cite{VN19}.
Moreover, the study of magnetic field effects is omitted due to the very simple dependence (linear)
of $\Omega_{\rm R}$ on the strength of $B_x$ for reasonably strong fields up to 1T.

\begin{figure}
\centering
\includegraphics[width=8.5cm]{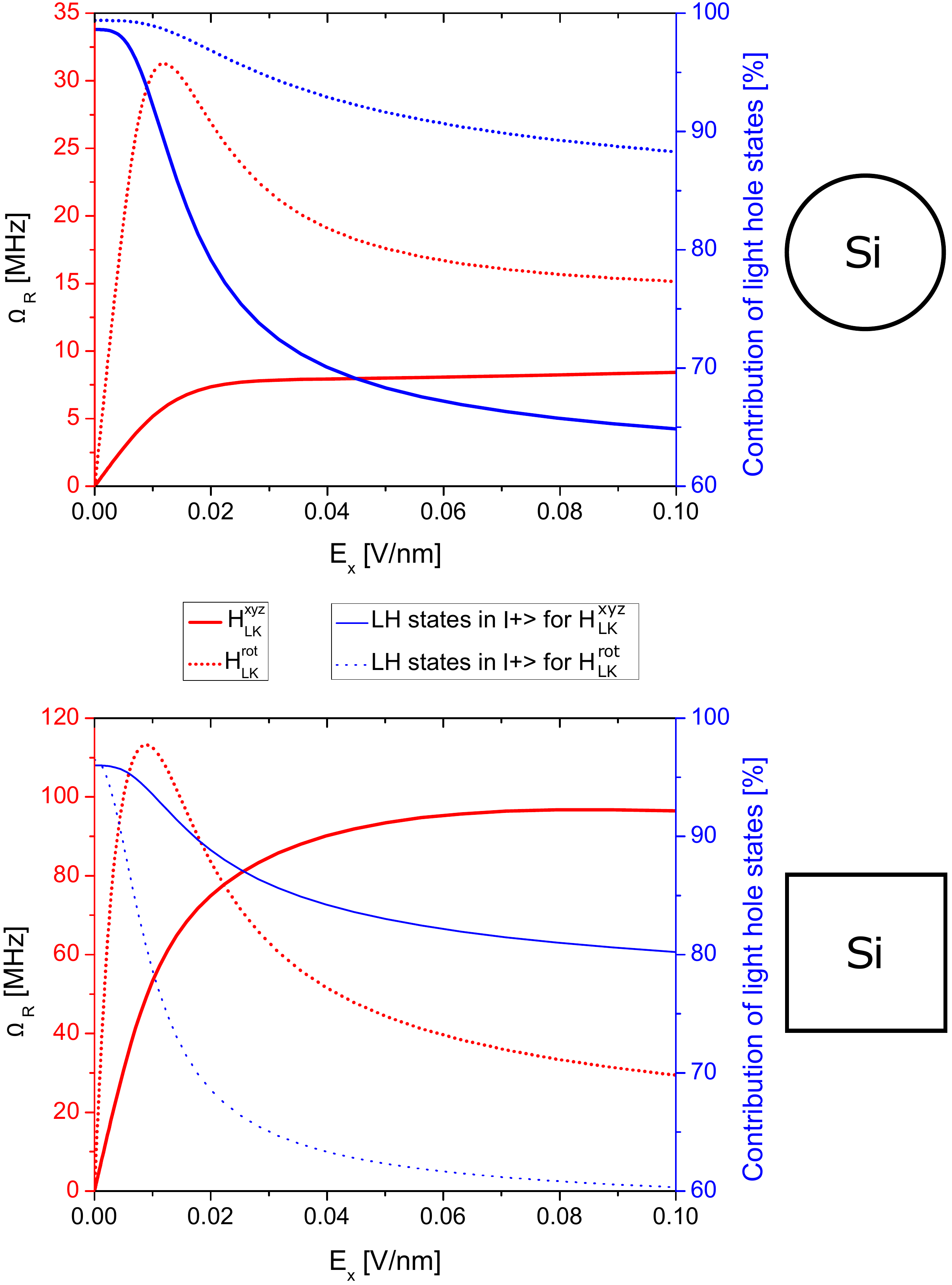}
\caption{Rabi frequency $\Omega_{\rm R}$ dependence on the electric field 
${\bf E}=(E_x,0,0)$ strength for two different nanowire orientations,
assuming circular (upper panel) and square (lower panel)
profile of the corresponding nanowire. 
The magnetic field applied is equal to $B=(0.1,0,0)$T, while other parameters can be found in the main text.
For all the configurations studied, the contribution of the light-hole states in the qubit state $\ket{+}$ versus the electric field strength is also given.}
\label{SiEDSR}
\end{figure}

In FIG.~\ref{SiEDSR}, the dependence of $\Omega_{\rm R}$ on the electric field strength $E_x$ is plotted for two different orientations of the nanowire and different cross sections, assuming $B_x=0.1$T. For weak electric fields, the Rabi frequency is linearly dependent on the electric field strength. A signature of the host Si material is the fact that $\Omega_{\rm R}$ differs significantly for different nanowire orientations, as expected since $\gamma_3/\gamma_2\gg1$. The LK Hamiltonian
$H_{\rm LK}^{\rm rot}$ corresponds to the orientation where the spin-orbit effects are the most pronounced~\cite{KRL18}; as a comparison, the Rabi frequency dependence on $E_x$ for the nanowire orientation that coincides with the main crystallographic axes is given.  Also, in FIG.~\ref{SiEDSR}, the percentage of light-hole states ($j_z=\pm1/2$) in the qubit state $\ket{+}$ (very similar result is obtained for $\ket{-}$) is given for different cross sections and nanowire orientations. For weak electric fields, $\ket{+}$ is almost exclusively of the light-hole origin, as suggested when analyzing Si nanowire states at the $\Gamma$ point~\cite{KRL18}. However, for stronger fields, the influence of heavy-hole states enhances and can become significant as the electric field strength is further increased. 

The response of the Rabi frequency to the applied electric field can be divided into two regimes: the first one corresponds to the linear response of $\Omega_{\rm R}$, whereas in the second regime nonlinear response is present. 
In the linear regime $\Omega_{\rm R}$ can be very well approximated as $\alpha_{\rm sub}^{\rm sup}E_x$, $\Omega_{\rm R}\approx\alpha_{\rm sub}^{\rm sup}E_x$, where the subscript/superscript corresponds to the nanowire profile/orientation.
The corresponding parameters $\alpha_{\rm s}^{\rm xyz}=6.40\times10^{-6}\,{\rm MHz\, \frac{m}V}$,  $\alpha_{\rm s}^{\rm rot}=3.26\times10^{-5}\,{\rm MHz\, \frac{m}V}$, 
$\alpha_{\rm c}^{\rm xyz}=5.91\times10^{-7}\,{\rm MHz\, \frac{m}V}$, and
$\alpha_{\rm c}^{\rm rot}=4.11\times10^{-6}\,{\rm MHz\, \frac{m}V}$ indicate that the square profile is by far more suitable for the electrical control of the hole spin qubit, as well as $z\,||$\,[001], $x\,||$\,[110] orientation.
A more complicated dependence on $E_x$ is observed in the nonlinear regime: 
for both the square and circular profile having the $z\,||$\,[001], $x\,||$\,[110] orientation, global maximum of $\Omega_{\rm R}$ around $10^{-2}\,{\rm V/nm}$ is observed, followed by the rapid decline of Rabi frequency. On the other hand, for the $xyz$ orientation and both the sNW and cNW QD, $\Omega_{\rm R}$ is weakly dependent on the electric field strength.  

The orbital contribution analysis in each scenario can be used to rationalize the obtained results. Besides the total heavy-hole/light-hole separation that was already given in FIG.~\ref{SiEDSR}, it is possible to match the orbital contribution of heavy-hole/light-hole basis states in qubit states $\ket{+/-}$. For weak electric fields, an orbital ground state with light-hole spins is dominant ($\ket{11}$ and $\ket{111}$ for Si cNW and sNW QD, respectively;
the formal definition of orbital states $\ket{in_z}$ and $\ket{n_xn_yn_z}$ can be found in Sec.~\ref{dijagonalizacija} and more specifically in Eqs.~\eqref{eigen} and~\eqref{eigen-circle}). For stronger fields, other states start to appear as well. 
Since the electric field is applied in the $x$-direction, it is reasonable to expect that higher orbital states in the $x$-direction appear in the decomposition of the basis states.  As we will show, for weak and moderate electric fields, only a few lowest-orbital states with heavy-hole and light-hole spins significantly affect the qubit states. Thus, we will demonstrate that it is possible to approximately describe the Rabi frequency in terms of only a few basis states. In the simplest picture, we can approximate the qubit states as
\begin{eqnarray}
  &&\ket{+/-}^{\rm c}\approx c_{11\pm\frac{1}2}^{\ket{+/-}}\ket{11}\ket{\pm\frac{1}2}+
  c_{12\pm\frac{1}2}^{\ket{+/-}}\ket{12}\ket{\pm\frac{1}2},\label{pmCIRC}\\
  &&\ket{+/-}^{\rm s}\approx s_{111\pm\frac{1}2}^{\ket{+/-}}\ket{111}\ket{\pm\frac{1}2}+
  s_{112\pm\frac{1}2}^{\ket{+/-}}\ket{112}\ket{\pm\frac{1}2},\label{pmSQUARE}
\end{eqnarray}
where $c_{11\pm\frac{1}2}^{\ket{+/-}}$, $c_{12\pm\frac{1}2}^{\ket{+/-}}$ 
($s_{111\pm\frac{1}2}^{\ket{+/-}}$, $s_{112\pm\frac{1}2}^{\ket{+/-}}$) represent coefficients of the qubit states $\ket{+}$ and $\ket{-}$ in Si cNW (sNW) QD.
It is important to mention that  the nonzero orbital contribution of states $\ket{12}\ket{\pm\frac{1}2}$ ($\ket{112}\ket{\pm\frac{1}2}$) is necessary to achieve $\Omega_{\rm R}\neq0$. This stems from the fact that the matrix element of $z$, appearing in the definition of $\Omega_{\rm R}$~\eqref{RabiFreqDef}, 
is zero if the $z$-component of the qubit states is the same, $\bra{1}z\ket{1}=0$, due to symmetry. Since $|\bra{1}z\ket{i}|$ is the largest for $i=2$, there follows the reason 
for choosing exactly this state in the Eqs.~\ref{pmCIRC} and~\ref{pmSQUARE}. 
\begin{figure}
\centering
\includegraphics[width=8.5cm]{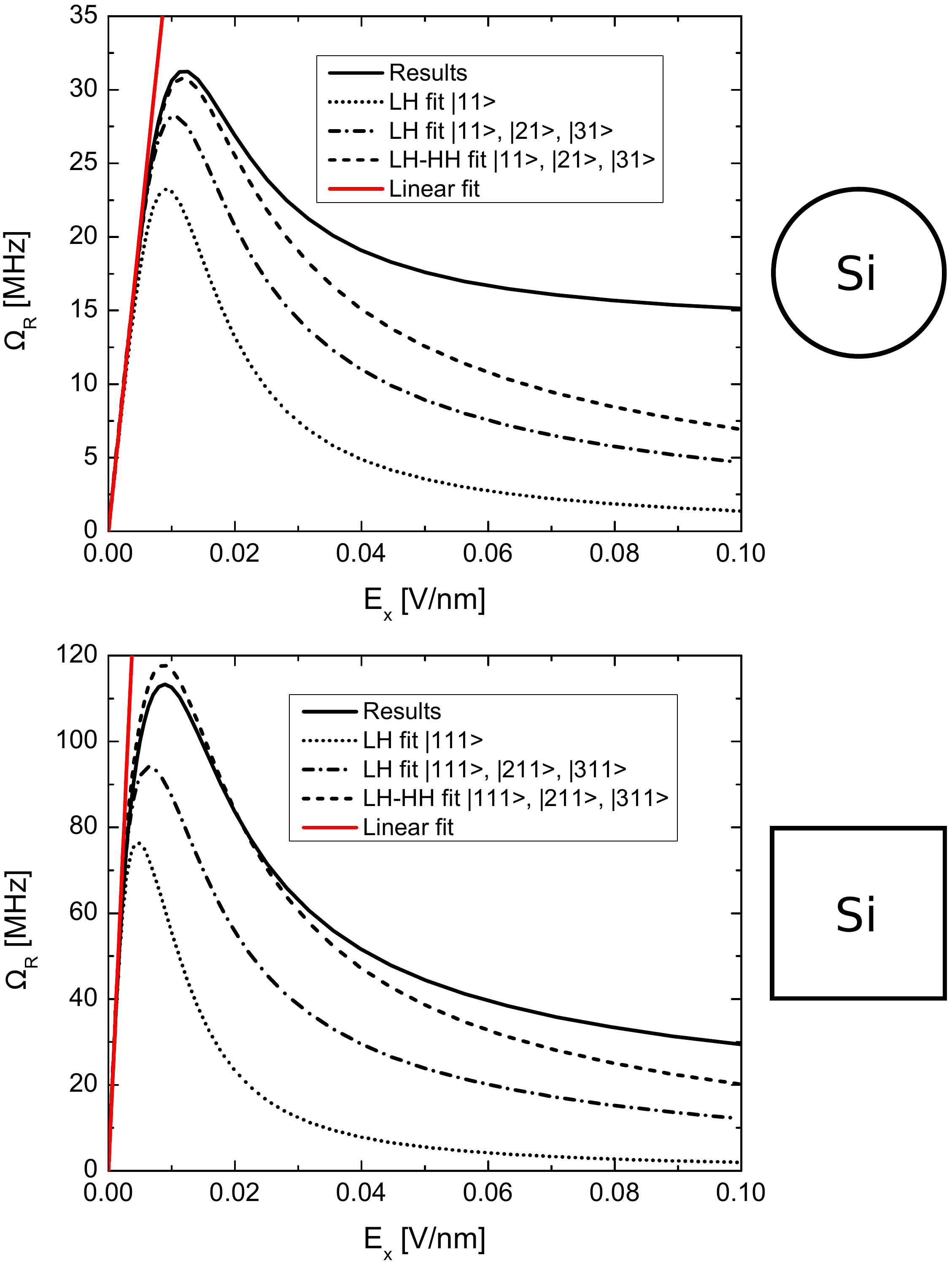}
\caption{For Si nanowire with circular (upper panel) and square (lower panel) cross section, described by the LK Hamiltonian $H_{\rm LK}^{\rm rot}$, comparison between the numerical results, linear fit of $\Omega_{\rm R}$, and different approximations of $\Omega_{\rm R}$ values are given: dotted line represents results
when qubit states are approximated with $\ket{11}$ and $\ket{12}$ (LH fit $\ket{11}$) light-hole states for cNW QD or with $\ket{111}$ and $\ket{112}$ (LH fit $\ket{111}$) light-hole states for sNW QD; dashed-dotted line/dashed line represent results
when states $\ket{11}$, $\ket{21}$, $\ket{31}$, $\ket{12}$, $\ket{22}$, $\ket{32}$  ($\ket{111}$, $\ket{211}$, $\ket{311}$, $\ket{112}$, $\ket{212}$, $\ket{312}$) having light-hole/light-hole and heavy-hole spins are used to approximate the qubit states.}
\label{SiRabiFIT}
\end{figure}
In FIG.~\ref{SiRabiFIT}, for the case of the LK Hamiltonian $H_{\rm LK}^{\rm rot}$, a comparison between the numerical results and approximations  at various levels are given. 
In the simplest case, we use the lowest orbital eigenstate and its pair to approximately describe the qubit states (dotted line). Also, we analyze the approximation with added orbital states $\ket{21}$, $\ket{31}$, $\ket{22}$, $\ket{32}$ ($\ket{211}$, $\ket{311}$, $\ket{212}$, $\ket{312}$) having light-hole spins (dashed-dotted line) and both the light-hole and heavy-hole spins (dashed line).
Finally, the linear fit of $\Omega_{\rm R}$ is plotted to determine the regime of linear response to the electric field. 

In the linear regime, it is evident that with only two orbital states, the behavior of $\Omega_{\rm R}$ can be explained. Since the coefficients   
$c_{11\pm\frac{1}2}^{\ket{+/-}}$ and $s_{111\pm\frac{1}2}^{\ket{+/-}}$
are independent on $E_x$ in this regime, the linear response to $E_x$ purely corresponds to the linear response of coefficients $c_{12\pm\frac{1}2}^{\ket{+/-}}$ and $s_{112\pm\frac{1}2}^{\ket{+/-}}$ to the applied electric field.
For stronger fields, the influence of the other orbital states and heavy-hole spins is evident (see FIG.~\ref{SiRabiFIT}), and it is necessary to expand the number of basis states for realistic approximation of $\Omega_{\rm R}$. The plots show that the increased number of basis states  gives more realistic results, thus explaining the need for the relatively large number of basis states included in the Hamiltonian diagonalization.

In the end, it remains an open question why the nanowire with a square profile is more susceptible to displaying large Rabi frequencies than the circular ones. To this end, we made an approximate model of the Rabi frequency based on the perturbation theory (see the Appendix~\ref{Perturbative} for more details). More concretely, we divide the total Hamiltonian~\eqref{hole-model} at finite ${\bf B}$ into two parts: the first one, $H_1$, corresponds to the case of zero magnetic field, while the second one, $H_2$, collects the magnetic-field-dependent terms and can be treated as a perturbation. Using the first-order perturbation theory, an approximate description of the Rabi frequency (see Eq.~\ref{PERTexpansion}) can be made, collecting the basic features of $\Omega_{\rm R}$. Numerical estimates, in this case, confirm that the main difference in Rabi frequency values of cNW and sNW QD hole spin qubits lies in much stronger magnetic-field-induced transition matrix elements of sNW QD, as explained in detail in the Appendix~\ref{Perturbative}.


\section{Hole spin qubit in Ge nanowire QD}\label{GeEDSRsection}

In this section, using the spherical approximation of the LK Hamiltonian, see Eq.~\ref{LKhamSPHERICAL}, the effect of nanowire geometry and the electric field strength on the Rabi frequency is going to be investigated. Again, we will discuss our results in the magnetic field regime up to 1T, for which the response of $\Omega_{\rm R}$ to $B_x$ is linear.
\begin{figure}
\centering
\includegraphics[width=7.6cm]{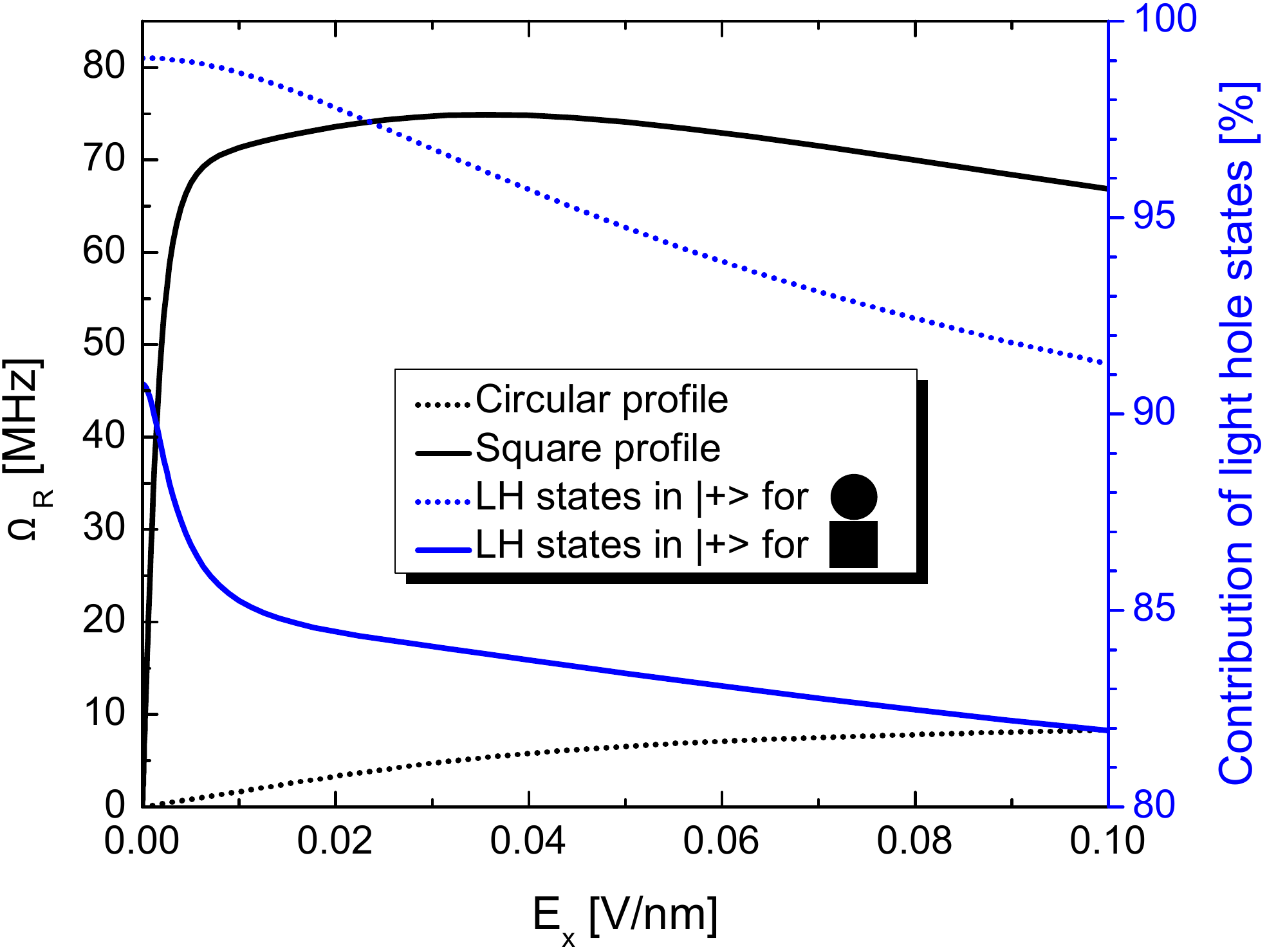}
\caption{For the hole spin qubit in Ge cNW and sNW QD, dependence of the Rabi frequency $\Omega_{\rm R}$ on the electric field strength $E_x$ is given. 
The applied magnetic field is equal to $B=(0.1,0,0)$T, while other parameters can be found in the main text. On the right side of the plot, for the configurations studied, the contribution of the light-hole states in the qubit state $\ket{+}$ versus the electric field strength is given.}
\label{GeEDSR}
\end{figure}

In FIG.~\ref{GeEDSR}, we plot the dependencies of $\Omega_{\rm R}$
on the electric field strength for the hole spin qubit in Ge cNW and sNW QD, assuming magnetic field strength $B_x=0.1$T. On the right side of the plot, we present the influence of light-hole states in $\ket{+}$  for the studied cases.

The obtained results can be divided according to the type of response to the electric field. In the linear regime, the fitting $\Omega_{\rm R,i}^{\rm Ge}=\alpha_{\rm i}^{\rm Ge} E_x$, $i=\rm s, c$, gives the parameters 
$\alpha_{\rm s}^{\rm Ge}=3.37\times10^{-5}\,{\rm MHz\, \frac{m}V}$ and $\alpha_{\rm c}^{\rm Ge}=1.62\times10^{-7}\,{\rm MHz\, \frac{m}V}$, confirming the beneficial role of the square profile for the electrical control of the hole spin qubit again. Comparing the results with the hole spin qubits inside the Si nanowire QD, the relation $\alpha_{\rm s}^{\rm Ge}\approx\alpha_{\rm s}^{\rm rot}$ indicates that both materials can be used to provide similar outputs. The nonlinear regime in Ge differs for different profiles. In the case of the sNW QD, Rabi frequency is weakly dependent on the electric field strength.
In contrast, $\Omega_{\rm R}$ in the cNW QD gradually increases.
However, due to the much stronger slope of $\Omega_{\rm R}$ in the linear regime, the hole spin qubit in Ge sNW QD gives much better results than the Ge cNW QD.

To gain better insight into the $\Omega_{\rm R}$ difference for Ge cNW and sNW QD hole spin qubit, we analyze the orbital contribution of qubit states in each case of interest.
First, we notice that the qubit state $\ket{+}$ in Ge cNW QD is almost exclusively of light-hole origin for weak electric fields (similar as in Si cNW and sNW QDs, see FIG.~\ref{SiEDSR}). In contrast, in Ge sNW QD, the influence of light-hole states is considerably smaller. Besides the different light-hole/heavy-hole influence, the orbital composition of qubit states in cNW and sNW QD differs greatly.
\begin{figure}
\centering
\includegraphics[width=8.5cm]{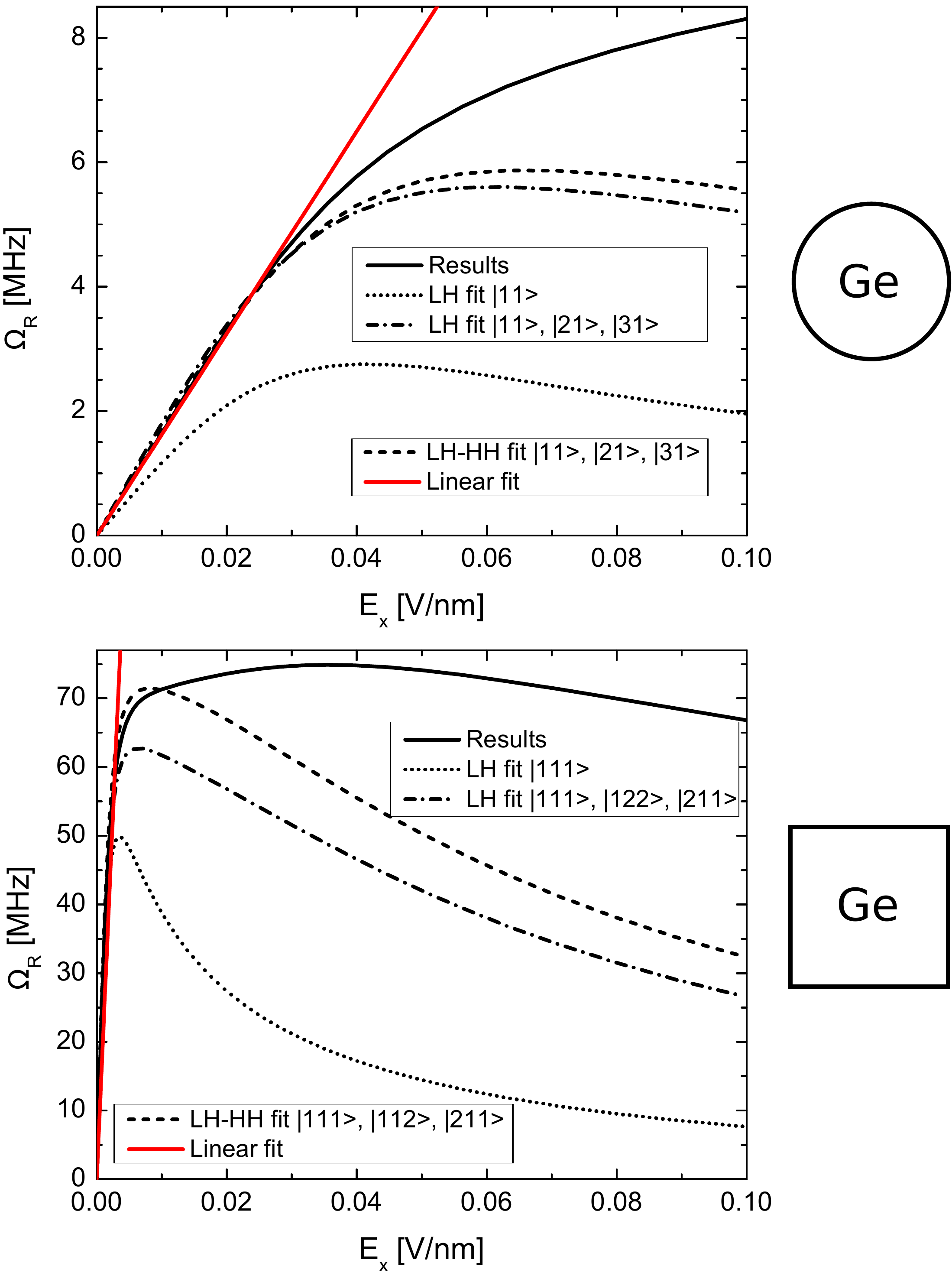}
\caption{For the hole spin qubits in Ge nanowire with circular (upper panel) and square (lower panel) cross section, comparison between the numerical results, the linear fit of $\Omega_{\rm R}$, and the fits of $\Omega_{\rm R}$ with approximated qubit states is given. In the case of Ge cNW QD, dotted line represent results when qubit states are approximated with orbital states $\ket{11}$ and $\ket{12}$ having light-hole spins; dashed-dotted/dashed line represent results
when states $\ket{11}$, $\ket{21}$, $\ket{31}$, $\ket{12}$, $\ket{22}$, $\ket{32}$  
having light-hole/light-hole and heavy-hole spins are used to approximate the qubit states. In the case of sNW QD, besides the linear fit and the numerical results, the minimal fit of qubit states (Eq.~\ref{pmSQUARE}) is used, as well as approximations based on states $\ket{111}$, $\ket{122}$, $\ket{211}$ with light-hole/light- and heavy-hole spins (see the last paragraph of Sec.~\ref{GeEDSRsection}).}
\label{GeRabiFIT}
\end{figure}
In the case of qubit states created in cNW QD, 
the role of the light-hole $\ket{11}$ state is dominant for electric field strengths up to $10^{-3}\,$V/nm. For stronger fields, the influence of orbital $\ket{21}$ and $\ket{31}$ states with both light-hole and heavy-hole spins becomes relevant. Similarly as for Si nanowires, we can approximate Rabi frequency behavior with only a few orbital states.
In FIG.~\ref{GeRabiFIT} we compare the obtained numerical results with the linear fit of $\Omega_{\rm R}$, the minimal model (LH fit $\ket{11}$) that approximates  qubit states with light-hole states $\ket{11}$ and $\ket{12}$, as well as
approximations taking into account orbital states $\ket{11}$, $\ket{12}$, $\ket{21}$, $\ket{22}$, $\ket{31}$, and $\ket{32}$ with: i) light-hole (LH fit $\ket{11}$, $\ket{21}$, $\ket{31}$), ii) both the light-hole and heavy-hole spin states (LH-HH fit $\ket{11}$, $\ket{21}$, $\ket{31}$).
As evident from the upper panel of FIG.~\ref{GeRabiFIT}, the minimal model can reproduce the linear response of $\Omega_{\rm R}$; for stronger fields, more states are needed to reproduce the Rabi frequency results adequately. 

The orbital contribution of Ge sNW QD qubit states is significantly different. For weak electric fields, up to $10^{-4}\,$V/nm, besides the light-hole state $\ket{111}$, which is relatively dominant, several other states appear as well. To mention a few, light-hole states $\ket{122}$, and $\ket{211}$ are present with both light- and heavy-hole spin states. In the minimal basis model, see Eq.~\ref{pmSQUARE}, it is possible to qualitatively explain the behavior of $\Omega_{\rm R}$ for weak fields. For stronger electric fields, qubits have more and more disperse orbital contribution, and more basis states need to be included in the picture. This is illustrated in the lower panel of  FIG.~\ref{GeRabiFIT}, where, besides the linear and the minimal basis fit (LH fit $\ket{111}$), approximation based on the orbital states $\ket{111}$, $\ket{122}$, $\ket{211}$ with
light-hole/light- and heavy-hole spins are presented (LH/LH-HH fit $\ket{111}$, $\ket{122}$, $\ket{211}$). Note that in the case of $\ket{111}$ and $\ket{211}$ states, complementary states $\ket{112}$ and $\ket{212}$ are included in the picture because $|\bra{1}z\ket{2}|$ is the dominant transition matrix element between those of the type $|\bra{1}z\ket{i}|$, $i=1,...,15$. On the other hand, for $\ket{122}$ both the $\ket{121}$ and $\ket{123}$ states should be included, since $|\bra{2}z\ket{1}|\approx|\bra{2}z\ket{3}|$.


\section{Hole spin qubit in Ge/Si core/shell nanowire QD}
\label{StrainSection}
Finally, we study the strain effects in Ge/Si core/shell nanowire QD hole spin qubit. Since the Bir-Pikus Hamiltonian  is proportional to
$J_z^2$, it can be concluded that
the (degenerate) eigenstates of $H_{\rm BP}^{\rm eff}$ are either light-hole states $\chi_{\pm1/2}$ with the eigenvalue $|b|(\epsilon_{\perp}(\gamma)-\epsilon_{zz}(\gamma))/4$
or heavy-hole states $\chi_{\pm3/2}$ with the eigenvalue $9|b|(\epsilon_{\perp}(\gamma)-\epsilon_{zz}(\gamma))/4$. 
Focusing on the qubit states, being the two lowest eigenstates of the total Hamiltonian, it is evident that the strain will lead to the increase of the light-hole/heavy-hole contribution in qubit states, depending on the sign and value of $\epsilon_{\perp}(\gamma)-\epsilon_{zz}(\gamma)$, where $\gamma$ represents the relative shell thickness. The value of $\epsilon_{\perp}(\gamma)-\epsilon_{zz}(\gamma)$ is always positive and increases with the $\gamma$ increase. Thus, we conclude that the role of light-hole states in $\ket{+/-}$ will be enhanced.
\begin{figure}
\centering
\includegraphics[width=7cm]{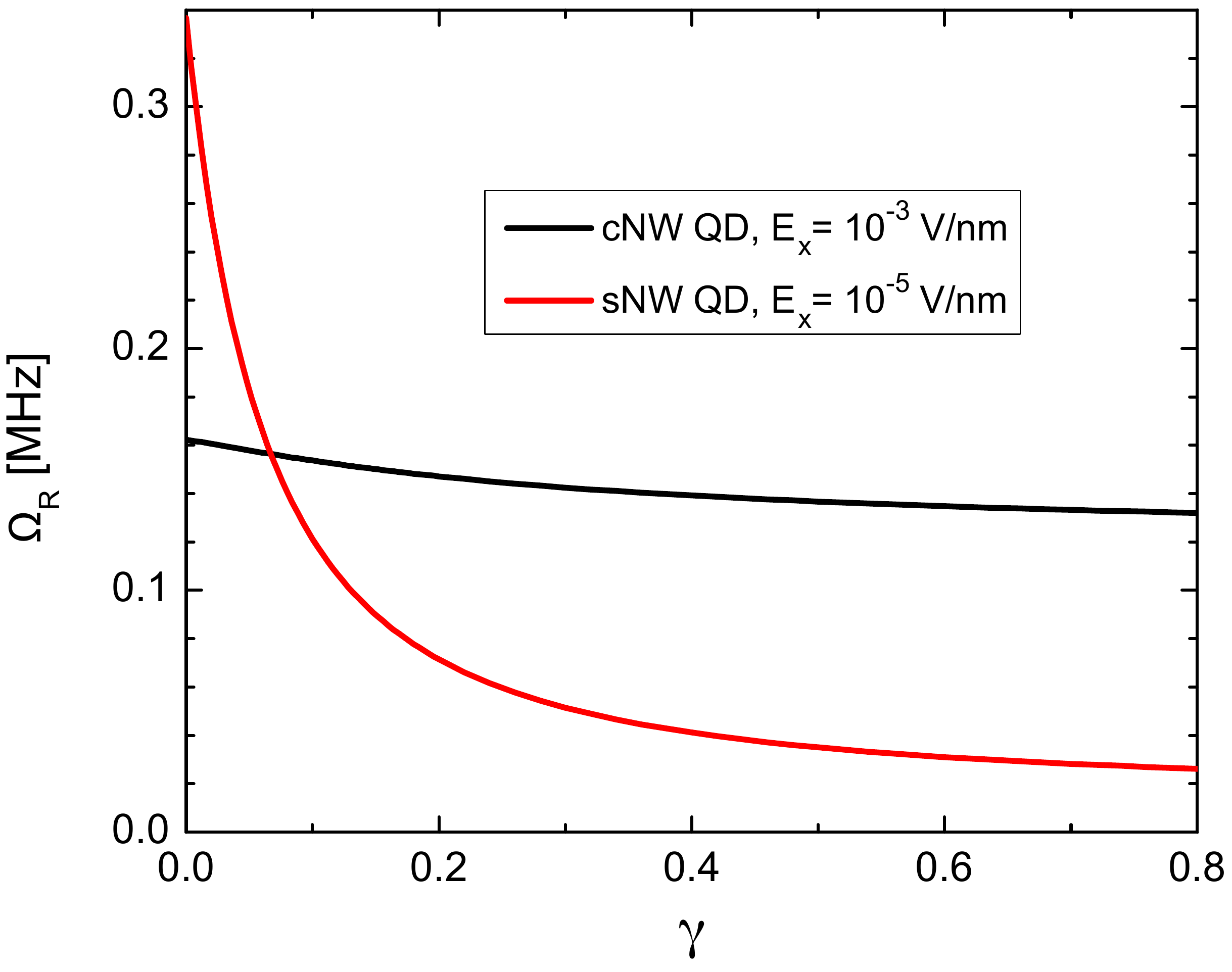}
\caption{Dependence of $\Omega_{\rm R}$ on strain in hole spin qubits formed inside Ge/Si core/shell cNW and sNW QD. 
In both cases, magnetic field $B_x$ is equal to 0.1T, whereas the electric field values are set to $10^{-3}\,$V/nm and 
$10^{-5}\,$V/nm for the circular and square profiles, respectively.}
\label{GeEDSRstrain}
\end{figure}

To study the response of the Rabi frequency on the presence of strain, we fix the  magnetic field strength to $0.1$T and vary both the electric field strength and shell thickness $\gamma\in(0,0.8)$. The results will be discussed for both the Rabi frequency's linear and nonlinear response to the electric field.  
In the linear regime, qubit states in cNW QD can be very well approximated as in Eq.~\eqref{pmCIRC}. The effect of strain is such that the orbital contribution of the light-hole state $\ket{11}$ slightly increases, whereas the light-hole state $\ket{12}$ 
decreases more rapidly; thus, it is expected for the Rabi frequency to decrease.
However, this decrease is weak, as evident from the fact that $\Omega_{\rm R}(\gamma=0.8)/\Omega_{\rm R}(\gamma=0)$ is approximately 1.2 for electric fields up to $10^{-3}\,$V/nm.  
On the other hand, in the Ge/Si core/shell sNW QD strain has a more pronounced effect, leading to the much stronger decreasing trend of $\Omega_{\rm R}$ with the increase of $\gamma$.
As an example, in FIG.~\ref{GeEDSRstrain}, the dependence of the Rabi frequency on the shell thickness $\gamma\in(0,0.8)$ is given for the hole spin qubit in Ge/Si core/shell cNW and sNW QD, assuming electric field strengths $10^{-3}\,$V/nm and $10^{-5}\,$V/nm, respectively.
It is evident from the plots that strain in sNW QD hole spin qubit leads to a one order of magnitude decrease of the Rabi frequency, whereas in cNW QD its effect is very weak.

Finally, a weak effect of strain on $\Omega_{\rm R}$ in Ge/Si core/shell cNW QD is preserved in the nonlinear regime also. On the other hand, in sNW QD
the role of strain is muffled with the increase of the electric field ($E_x\leq0.1\,$V/nm), reaching the ratio $\Omega_{\rm R}(\gamma=0.8)/\Omega_{\rm R}(\gamma=0)\approx 1.35$ that is comparable with nanowires having a circular profile. 
\section{Conclusions}\label{Conclusions}
We analyzed hole spin qubits in QD formed inside Si, Ge, and Ge/Si core/shell cNW and sNW.
The possibility to electrically control the hole spin qubit through direct Rashba spin-orbit coupling is exploited, and the role of different materials and geometries is investigated in detail, with the goal to find setups that enable the fastest control of the hole spin qubit. The Rabi frequency, the quantity that allows a simple estimation of the qubit efficiency, is defined, and its dependence on the electric field strength is investigated.
We showed that the hole spin qubits in QDs inside sNW are much more easily tuned than the corresponding qubits in cNW QDs.
For weak fields, the Raby frequency is linearly proportional to the electric field strength.
In this regime, we showed that the Rabi frequency in the hole spin qubits inside Ge and Si sNW QDs are of comparable strengths, providing that the orientation of Si nanowire is such that the spin-orbit effects are maximized. 
In the nonlinear regime, the global maximum of the Rabi frequency is found in QD inside Si sNW, putting this setup in favor of others for the creation of the hole spin qubit.
Finally, we studied strain effects in the hole spin qubit inside Ge/Si core/shell nanowire QD.
Our numerical analysis shows that strain diminishes the Rabi frequency. Whereas in cNW QD this effect is not so pronounced, a strong influence of strain in Ge/Si core/shell sNW QD is observed, such that it can be
optimized with the appropriate tuning of the electric field strength.

In the end, a few general remarks should be addressed.
First, our results are in line with the recent experimental work~\cite{FCM+2021}, assuming the same magnetic field strength.
Also, although our work was focused on the Rabi frequency between the two lowest (qubit) states, at nonzero temperature, due to thermal activation,
the electric-field-induced transition between different states can be achieved.
Although the detailed research is beyond the scope of this work,
it should be stated that the electrical control of the Rabi frequency is possible only with states that were initially degenerate at zero magnetic field, whereas between energetically divided states at $B=0$, Rabi frequency is only weakly sensitive on the strength of the electric field.
Also, in this work, we assumed hard-wall confinement.
According to the authors of~\cite{SM20}, a much stronger Rabi frequency is expected for
the smooth-wall confinement than in the setup studied within this work.
In future works, it would be interesting to investigate the role of
different regular shapes of nanotubes, having the symmetry ${\bf C}_{\rm nv}$,
where $n$ corresponds to the order of the rotational axis.
Since simpler regular shapes, such as equilateral triangle (n=3), 
hexagon (n=6), octagon (n=8) have a similar order of the rotational axis to the square profile (n=4) and much lower $n$ than the circular profile ($n=\infty$), it is to be expected that the results in such geometries should be more similar to the results for the square profile.
\acknowledgments
This research was funded by the Ministry of Education, Science, and Technological Development of the Republic of Serbia and the National Scholarship Programme of the Slovak Republic (ID 30281).
\appendix
\section{Rabi frequency based on first-order perturbation theory}\label{Perturbative}
Here, we present the estimate of the Rabi frequency based on the first-order perturbation theory, which can be used the gain more insight into the difference between the cNW and sNW QD hole spin qubits. 

Following the approach described in~\cite{VBP+18},
we divide the QD Hamiltonian $H$ into two parts: 
the first one $H_1$ collects the LK Hamiltonian (Eqs.~\ref{LKham} or \ref{LKhamROT}) at zero magnetic field, potential $V$, and DRSOI~\eqref{HDR},
\begin{equation}
    H_{1}=H_{\rm LK}^{\rm xyz/rot}({\bf B}=0)+V+H_{\rm DRSOI},
\end{equation}
while the second part, $H_2$, collects the magnetic-field-dependent parts consisting on the Zeeman term~\eqref{hh} and the magnetic field dependent part of the LK Hamiltonian
\begin{equation}
    H_{2}=H_{\rm Z}+(
    H_{\rm LK}^{\rm xyz/rot}({\bf B})-
    H_{\rm LK}^{\rm xyz/rot}({\bf B}=0)).
\end{equation}
In the zero field, the eigenvalues  $E_{n}$  of $H_1$,
with the corresponding eigenstates $\ket{n,\sigma=\uparrow\downarrow}$, are two-fold degenerate. We can define the corresponding qubit states $\ket{\pm_0}$ 
from the hole ground states $\ket{0,\pm}$ as an eigenvalues of the magnetic-field Hamiltonian $H_2$ (we neglect terms proportional to $B^2$, since the numerical results in the main text show that the Rabi frequency is linearly dependent on the magnetic-field strength),
\begin{equation}
\begin{pmatrix}
\bra{0,+}H_2\ket{0,+} & \bra{0,+}H_2\ket{0,-} \\
\bra{0,-}H_2\ket{0,+} & \bra{0,-}H_2\ket{0,-} 
\end{pmatrix}.
\end{equation}
Using the first-order correction of the qubit states
\begin{eqnarray}
\ket{+_1}&=&\ket{+_0}+\sum_{n\neq0,\sigma}
\frac{\bra{n,\sigma}H_2\ket{+_0}}{E_0-E_n},\\  
\ket{-_1}&=&\ket{-_0}+\sum_{n\neq0,\sigma}
\frac{\bra{n,\sigma}H_2\ket{-_0}}{E_0-E_n},
\end{eqnarray}
the Rabi frequency can be defined as 
\begin{eqnarray}\label{PERTexpansion}
    \Omega_{\rm R}=e\frac{E_{\rm nw}}{h}
    \Big|\sum_{n\neq0,\sigma}&&\frac{1}{E_0-E_n}
    \big(\bra{+_0}z\ket{n\sigma}\bra{n\sigma}H_2\ket{-_0}
    \nonumber\\
    &&+
    \bra{+_0}H_2\ket{n\sigma}\bra{n\sigma}z\ket{-_0}\big)\Big|.
\end{eqnarray}
Using the previous relation, the role of different contributions to the Rabi frequency can be determined:
the role of energy separation at zero $B$, the role of the dipole term $|\bra{\pm_0}z\ket{n\sigma}|$ and the role of the 
magnetic-field-induced transition matrix elements of the form
$|\bra{\pm_0}H_2\ket{n\sigma}|$. 
Also, it is instructive to look at the minimal number of states needed to satisfactorily describe the Rabi frequency. 

\begin{figure}[htp]
\centering
\includegraphics[width=6.4cm]{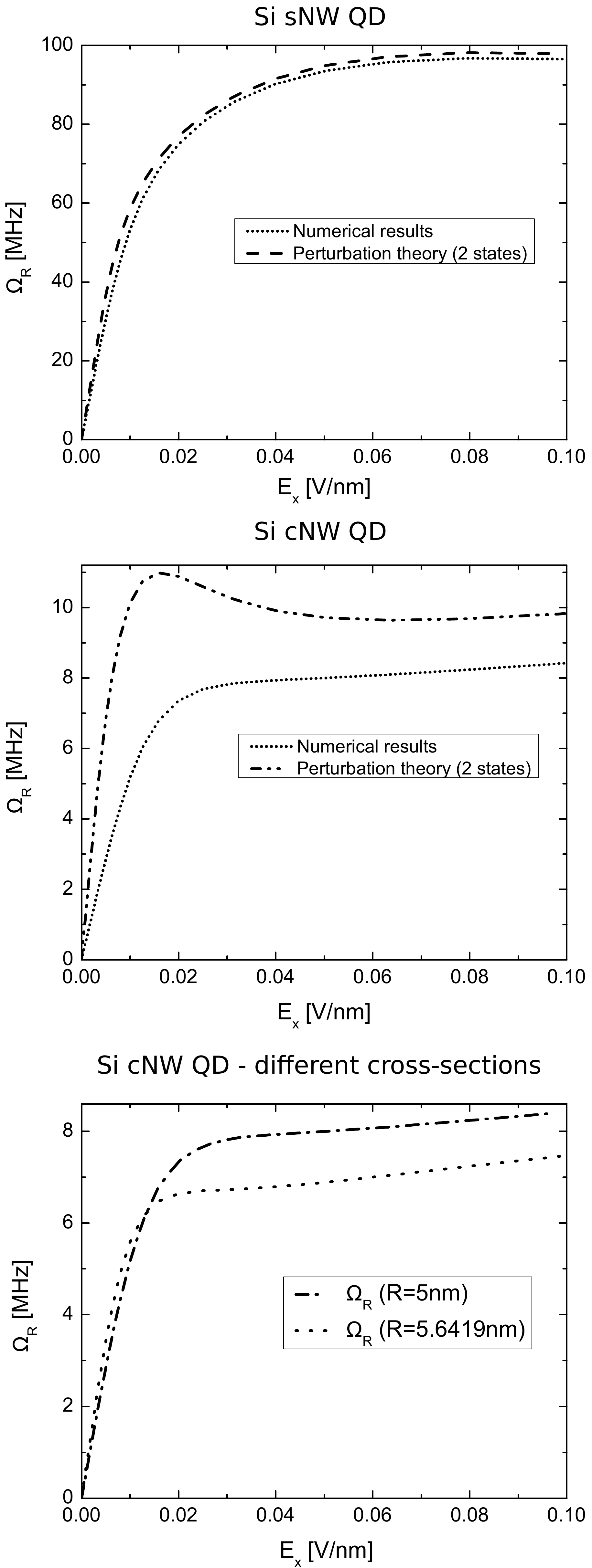}
\caption{(upper panel) In the case of the Si sNW QD hole spin qubit, a comparison between the numerical results and the approximation of $\Omega_{\rm R}$ using the perturbation theory (Eq.~\ref{PERTexpansion}) is given. Note that in the perturbative derivation of $\Omega_{\rm R}$ only two states that were initially degenerate at $B=0$ were included.
(middle panel) For the Si cNW QD hole spin qubit, 
comparison between the numerical results (dotted line) and the pertubative estimate of Rabi frequency results obtained
using two states in the pertubative expansion (dashed-dotted line) is given.
(lower panel) Comparison of $\Omega_{\rm R}$ for cNWs with half diameters 5~nm (used in all previous plots) and 5.6419~nm. 
In all calculations, we assumed that magnetic field $B_x$ is equal to 0.1~T and have used the LK Hamiltonian 
$H_{\rm LK}^{\rm xyz}$.}
\label{AppFIG}
\end{figure}
In the case of the sNW QDs only two states, having the same energy $E_1$ at $B=0$, should be included in the perturbative expansion to obtain very good approximation of the Rabi frequency. This is illustrated in the upper panel of FIG.~\ref{AppFIG} on the example of Si, using the LK Hamiltonian $H_{\rm LK}^{\rm xyz}$. 
With the help of the same LK Hamiltonian, in the middle panel of FIG.~\ref{AppFIG} we present the perturbative result in the case of Si cNW QD hole spin qubit, using the two states (dashed-dotted line) in the perturbative expansion, with the corresponding degenerate energy (at $B=0$) $E_1$. Although the fit is not ideal (we obtain the converged results by taking six states in the perturbative expansion), it can be used to qualitatively compare the role of the nanotube profile and to identify the term responsible for the significant difference between Rabi frequencies. 

To this end, we will compare the the perturbative results in the cNW and sNW case based on two-state perturbative expansion.
In the cNW case, the energy $E_1-E_0$ is weakly dependent on the electric field strength, varying the most 10$\%$ below 4.8 meV. The similar happens to the dipole term, which is almost constant and roughly equal to 3.7 nm. On the other hand, the magnetic-field-induced transition matrix element is strongly dependent on the electric field strength $E_x$ and, since the other two terms are weakly dependent on $E_x$, it follows almost the same dependence on the electric field as $\Omega_{\rm R}$.

The similar influence of the three terms can be traced in the sNW QD hole spin qubit also. The weak dependence of energy separation on $E_x$ is observed, changing at most 5$\%$ above/below 3 meV. In addition to that, the value of the dipole term was almost constant and proportional to 3.5 nm, while the magnetic-field-induced transition matrix element followed the electric field dependence of $\Omega_{\rm R}$.

Thus, much stronger Rabi frequency in the sNW QD case, when compared to the cNW QD, is due to much stronger 
magnetic-field-induced transition rate. The same conclusions are valid in the case of the LH Hamiltonian, $H_{\rm LK}^{\rm rot}$, confirming the dominant role of magnetic-field-induced transitions in determining the electric field dependence of $\Omega_{\rm R}$.

In the end, it remains an open question whether the different cross-sectional areas of the studied sNW and cNW can be responsible for obtaining different $\Omega_{\rm R}$. To this end, in the lower panel of FIG.~\ref{AppFIG} we present the results for cNWs with $R=5$ nm (old results) and $R=5.6419$ nm, where the last one corresponds to the same cross-sectional area as sNW. As is obvious from the plot, this difference is small, confirming that cross-section size plays no major role in the discrepancy between the two geometries. In the linear regime, $\Omega_{\rm R}(R=5.6419\,{\rm nm})>\Omega_{\rm R}(R=5\,{\rm nm})$, while for strong electric fields the opposite happens. This conclusion is consistent with~\cite{LLZ17}, where the Rashba hole effect is discussed for Si and Ge nanowires, being the dominant mechanism for the manipulation between the qubit states. 


\end{document}